\documentclass[twocolumn,linenumbers,times]{aastex631}
\usepackage{amsmath}
\usepackage[T1]{fontenc}

\newcommand{\menhance}{\dot{M}_{\mathrm{enh}}}

\begin{document}
\nolinenumbers
\title{Examining the Relationship Between the Persistent Emission and the Accretion Rate During a Type I X-ray Burst}

\author[0000-0001-9239-0373]{J. Speicher}\thanks{E-mail: jspeicher3@gatech.edu}\thanks{NDSEG Fellow}
\affiliation{Center for Relativistic Astrophysics, School of Physics, Georgia Institute of Technology, 837 State Street, Atlanta, GA 30332-0430, USA}

\author[0000-0001-8128-6976]{D. R. Ballantyne}
\affiliation{Center for Relativistic Astrophysics, School of Physics, Georgia Institute of Technology, 837 State Street, Atlanta, GA 30332-0430, USA}

\author[0000-0002-5786-186X]{P. C. Fragile}
\affiliation{Department of Physics \& Astronomy, College of Charleston, 66 George St., Charleston, SC 29424, USA}
% \affiliation{American Astronomical Society \\
% 1667 K Street NW, Suite 800 \\
% Washington, DC 20006, USA}

% \collaboration{20}{(AAS Journals Data Editors)}

% \author{F.X Timmes}
% \affiliation{Arizona State University}
% \affiliation{AAS Journals Associate Editor-in-Chief}

% \author{Amy Hendrickson}
% \altaffiliation{AASTeX v6+ programmer}
% \affiliation{TeXnology Inc.}

% \author{Julie Steffen}
% \affiliation{AAS Director of Publishing}
% \affiliation{American Astronomical Society \\
% 1667 K Street NW, Suite 800 \\
% Washington, DC 20006, USA}



\begin{abstract}
The accretion flow onto a neutron star will be impacted due to irradiation by a Type I X-ray burst. The burst radiation exerts Poynting-Robertson (PR) drag on the accretion disk, leading to an enhanced mass accretion rate. Observations of X-ray bursts often find evidence that the normalization of the disk-generated persistent emission (commonly denoted by the factor $f_a$) increases during a burst, and changes in $f_a$ have been used to infer the evolution in the mass accretion rate due to PR drag. Here, we examine this proposed relationship between $f_a$ and mass accretion rate enhancement using time-resolved data from simulations of accretion disks impacted by Type I X-ray bursts. We consider bursts from both spinning and non-spinning neutron stars and track both the change in accretion rate due to PR grad and the disk emission spectra during the burst. Regardless of the neutron star spin, we find that $f_a$ strongly correlates with the disk temperature and only weakly follows the mass accretion rate (the Pearson correlation coefficients are $\leq 0.63$ in the latter case). Additionally, heating causes the disk to emit at higher energies, reducing its contribution to a soft excess. We conclude that $f_a$ cannot accurately capture the mass accretion rate enhancement and is rather a tracer of the disk temperature.%especially when properties of the neutron star system could alter the relationship further. 
\end{abstract}


\section{Introduction } \label{sec:intro}
% \begin{itemize}
%     \item what are X-ray bursts?
%     \item X-ray burst change persistent emission
%     \item \cite{Worpel2013ApJ...772...94W,Worpel2015ApJ...801...60W}: scaling factor $f_a$
%     \item $f_a$ common tool in X-ray burst analysis, used to account for soft excess 
%     \citep[e.g.][]{Keek2018ApJ...855L...4K,Bult2019ApJ...885L...1B,Guver2022MNRAS.510.1577G, Guver2022ApJ...935..154G}
%     \item suggestion: $f_a$ increasing due to mass accretion rate enhancement due to PR drag
%     \item \cite{Fragile2018ApJ...867L..28F,Fragile2020NatAs...4..541F,Speicher2023MNRAS.526.1388S} found that $\dot{M}$ enhanced
%     \item \cite{Fragile2020NatAs...4..541F,Speicher2023MNRAS.526.1388S}: temperature increase too due to viscous heating$\rightarrow$ persistent emission increases
%     \item tracing $\dot{M}$ with $f_a$
% \end{itemize}

Neutron stars in low-mass X-ray binaries accrete matter from their companion star via Roche-Lobe overflow. The accreted matter moves inwards through an accretion disk aided by viscous processes moving angular momentum outwards \citep[for a review, see][]{Abramowicz2013LRR....16....1A}. The accretion liberates potential energy, a large fraction of which is radiated by the disk. %The accretion disk dissipates half of the total accretion luminosity
%the rest at the neutron star or a boundary layer. 
%and radiates a fraction of this liberated energy away \pcf{I don't really understand this sentence. Do you really dissipate a luminosity? And if it only radiates a fraction of that luminosity, what happens to the rest?}. 
If the accretion disk is optically thick, its spectrum is expected to be consistent with a multicolor blackbody, where the blackbody temperature varies with radius \citep[e.g.,][]{Mitsuda1984PASJ...36..741M,Kubota2005ApJ...631.1062K}.% ,Bult2019ApJ...885L...1B,Lu2024arXiv240414129L}.

% The accretion disk emission depends on the disk properties such as temperature and will change when impacted by a 
The accretion disk around neutron stars in these systems can be impacted by Type I X-ray bursts \citep[for a review of Type I X-ray bursts, see ][]{Galloway2021ASSL..461..209G}. Type I X-ray bursts occur due to the ongoing accumulation of accreted material onto the neutron star surface. The accumulation of matter gradually increases the temperature and pressure at the neutron star surface, which can cause unstable nuclear burning. This unstable burning consumes the accumulated matter and heats the neutron star. The resulting X-ray burst radiates $\sim10^{39}$ ergs away over a time span usually lasting on the order of tens of seconds \citep[e.g.,][]{Bult2019ApJ...885L...1B,Galloway2020ApJS..249...32G,Guver2022ApJ...935..154G,Yan2024MNRAS.529.1585Y}. 

The burst radiation illuminates the neutron star environment, leading to structural and spectral changes of the accretion flow \citep[for a review, see][]{Degenaar2018SSRv..214...15D}. A common observation is a drop in hard X-ray emission during bursts \citep[e.g.][]{Maccarone2003A&A...399.1151M,Ji2014ApJ...782...40J,SanchezFernandez2020A&A...634A..58S,Peng2024A&A...685A..71P}, which has been connected to the cooling of the corona, a region containing hot electrons \citep[e.g.,][]{Fragile2018ApJ...867L..28F,Speicher2020MNRAS.499.4479S}. Reprocessed emission from the accretion disk produces a reflection spectrum. During X-ray bursts, reflection is observable, for instance, through the detection of emission lines \citep[e.g.,][]{Ballantyne2004ApJ...602L.105B,Degenaar2013ApJ...767L..37D,Keek2017ApJ...836..111K,Bult2019ApJ...885L...1B}, which will evolve during bursts \citep{Speicher2022MNRAS.509.1736S}. 

Furthermore, observations suggest that the burst enhances the accretion flow emission. \cite{Worpel2013ApJ...772...94W} and \cite{Worpel2015ApJ...801...60W} introduced a normalization factor $f_a$ for the accretion-powered preburst emission, also called the persistent emission, and showed that $f_a>1$ during most bursts. Since its introduction, $f_a$ has become a common tool in burst analysis due to its ability to improve spectral fits \citep[][]{Jaisawal2019ApJ...883...61J,Bult2022ApJ...940...81B,Yu2024A&A...683A..93Y,Lu2024ApJ...969...15L}, and also by being able to account for a soft excess below 3 keV \citep[e.g.,][]{Keek2018ApJ...855L...4K,Chen2022ApJ...936...46C,Guver2022ApJ...935..154G}.

The increase in the persistent emission during the burst has been attributed to Poynting-Robertson (PR) drag \citep{Robertson1937MNRAS..97..423R}. Burst photons exert PR drag onto the disk material, removing angular momentum. The angular momentum loss expedites the material infall and thus increases the mass accretion rate, which raises the accretion luminosity and its emission \citep{Walker1989ApJ...346..844W,Walker1992ApJ...385..642W}. Simulations of accretion flows impacted by X-ray bursts showed that PR drag can enhance the mass accretion rate by over an order of magnitude \citep{Fragile2018ApJ...867L..28F,Fragile2020NatAs...4..541F,Speicher2023MNRAS.526.1388S}, similar to the increase in $f_a$ inferred from burst observations \citep[e.g.,][]{Worpel2015ApJ...801...60W,Bult2022ApJ...940...81B}.

However, how the mass accretion rate enhancement translates into $f_a$ is not understood. While in the thin disk simulations by \cite{Fragile2020NatAs...4..541F} and \cite{Speicher2023MNRAS.526.1388S} the mass accretion rate significantly increased during the burst, the burst irradiation also heated the accretion disk. The increase in temperature at the accretion disk surface will modify its spectrum, and the heating effect could be mistaken as the impact of an enhanced mass accretion rate. Likewise, while $f_a$ can be used to fit the soft excess, this feature can also be explained with reflection \citep[e.g.][]{Bult2019ApJ...885L...1B,Speicher2022MNRAS.509.1736S,Zhao2022A&A...660A..31Z}.

This paper examines the relationship between the normalization factor $f_a$ and the mass accretion rate enhancement using the simulation data by \cite{Speicher2023MNRAS.526.1388S}. In addition, we will investigate the suitability of $f_a$ to account for the soft excess. In Section \ref{sec:methods}, we outline our calculations, with results presented in Section \ref{sec:results}. We discuss the results in Section \ref{sec:discussion} and conclude with Section \ref{sec:conclusion}. 
\section{Methods}\label{sec:methods}
\subsection{Calculation of the disk emission}\label{subsec:methodCalculationEmission}

We calculate the emission of the accretion disk using data from the four \textit{Cosmos++} \citep[e.g.,][]{Anninos2005ApJ...635..723A,Fragile2012ApJS..201....9F,Fragile2014ApJ...796...22F} simulations analyzed by \cite{Speicher2023MNRAS.526.1388S}. The simulation space covers a radial range of 10.7 km $\lesssim r\lesssim$350 km and a polar angular range of $0\lesssim \theta\lesssim\pi$. The space is subdivided into cells, with more being concentrated towards the equatorial plane and smaller radii. Each simulation consists of a neutron star within the inner simulation boundary and an accretion disk within the simulation space. The neutron star has a mass of $M=1.45$ M$_\odot$ and radius $R=10.7$ km. In two simulations, the neutron star does not spin ($a_*=0$), while in the other two simulations, $a_*=0.2$ and the rotational frequency is 500 Hz. The spin parameter $a_*$ is calculated as $a_*=cJ/GM^2$, where $J$ is the angular momentum \citep[for further details, see ][]{Speicher2023MNRAS.526.1388S}. 

The accretion disks are initialized following the gas-pressure-dominated regime of the $\alpha$ model \citep{Shakura1973A&A....24..337S}, with the viscosity parameter $\alpha=0.025$. Initially, the disks are in hydrostatic and thermal equilibrium. We assume that all magnetic fields are weak enough to be negligible.

\begin{figure}
    \centering
    \includegraphics[width = 0.45 \textwidth]{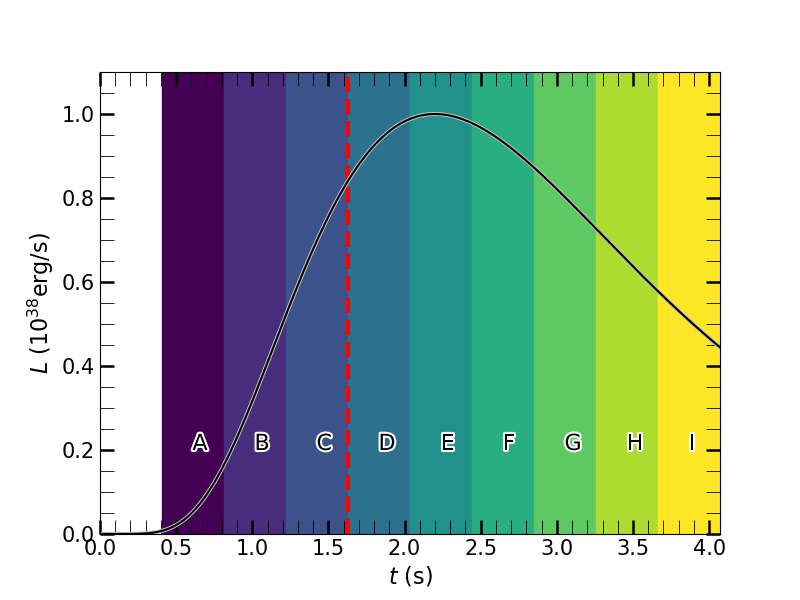}
    \caption{Luminosity profile (eq. \ref{eq:luminosity}) of the burst in the simulations by \cite{Speicher2023MNRAS.526.1388S}. The luminosity (black solid line) peaks at $10^{38}$ erg s\textsuperscript{-1} at $\approx2.2$ s. The shaded regions mark $\approx0.4$s long time intervals, during which we time-average the simulation data to calculate the disk emission (eq. \ref{eq:nuFnu}). Due to a transient numerical event, we only analyze the time intervals past the red dashed line for the $a_*=0.2$ burst simulation. For all no-burst simulations we use time interval F to calculate disk properties. }
    \label{fig:lightcurve}
\end{figure}

 %In  The data is time-averaged over $\approx0.4$ s time intervals, starting at $\approx0.4$ s. The shaded regions in Fig.\ref{fig:lightcurve} mark the time intervals with respect to the lightcurve of the simulated burst (black solid line). In the simulation, 
In each simulation set, the neutron star is either quiescent or emits burst radiation isotropically as observed in the neutron star's rest frame. The luminosity $L$ of the simulated bursts (black solid line in Fig.\ref{fig:lightcurve}) follows the lightcurve \citep[e.g.,][]{Norris2005ApJ...627..324N,Barriere2015ApJ...799..123B},
\begin{equation}
    L(t) = L_0 e^{2\left ( \tau_1/\tau_2 \right )^{1/2}} e^{-\frac{\tau_1}{t-t_s}-\frac{t-t_\mathrm{s}}{\tau_2}},\label{eq:luminosity}
\end{equation}
where $L_0=10^{38}$ erg s$^{-1}$, $t_\mathrm{s} = -0.25$ s, $\tau_1=6$ s, and $\tau_2=1$ s. Within the simulation space, \textit{Cosmos++} solves for the interaction between the burst radiation and the gas of the disk by evolving general relativistic, radiative,
viscous hydrodynamics equations \citep[e.g.,][]{Fragile2018Instability}.  %for further details regarding the simulation setup, see also \cite{Speicher2023MNRAS.526.1388S}

We time-average gas-temperature and mass density data from the simulation within several $\approx0.4$ s long time intervals (Fig. \ref{fig:lightcurve}, shaded areas) with each time interval separated by $\approx7\times10^{-3}$ s. 
%\pcf{What is 0.07 s? Is that the time interval between dumps? It's definitely not the simulation timestep, which would have been much smaller.}. 
The burst simulations ran for $\approx4.1$ s, yielding nine time intervals labeled A through I. The no-burst simulations ran for $\approx3.2$ s, but because disk properties vary negligibly in this setup we analyze data from their time-interval F. 

During each time interval, we assume that the disk radiates as a multicolor blackbody. Its emission is evaluated between an inner radius $r_{in}$ and an outer radius $r_{out}$, \citep[e.g.,][]{Zdziarski2022ApJ...939L...2Z}, 
\begin{equation}
  \nu L_\nu = \frac{4\pi h\nu^4}{c^2}\\
  \int_{r_{in}}^{r_{out}} \frac{r\,ds}{f_{c}\left [ T_{eff} \right ]^4 \left\{ \exp\left [ h\nu/kf_c T_{eff} \right ]-1\right\}},  \label{eq:nuFnu}
\end{equation}
where $h$ is Planck's constant, $c$ the speed of light and $T_{eff}$ the effective temperature. As the disk is not flat and changes shape during the simulations with bursts the incremental length $ds= \sqrt{1+\left(\frac{dz}{dr}\right)^2}dr$. We fix $dr=10$~m and interpolate the simulation data to find $r$ and $z$ for use in Eq.~\ref{eq:nuFnu}.
%depends on the incremental radial change $dr=10$ m and the incremental height change $dz$. Because the spatial simulation coordinates are not linearly spaced and $dr$ is smaller than the spatial simulation resolution, we need to interpolate both $r$ and $z$ to calculate an appropriate $dz$ for each increment in $dr$. We choose the linear interpolation function from \texttt{numpy} \citep{harris2020array}. %\pcf{I don't know what you mean you interpolate $r$ with $z$.}.
%\jsp{I interpolate the data because the radial spacing is not uniform. I tried to avoid interpolation by integrating using log(r). However, if I wanted to use the simulation data directly, I need to specify the radial shell corresponding to the inner radius. The shell can be chosen in different ways, and the figure in the pictures folder called 'spectraLogR\_options.png' shows that already a slight change in $r_{in}$ changes the spectrum (because the temperatures can be very high near $r_{in}$). It might therefore be the easiest to keep the interpolation.}

The inner radius $r_{in}$ corresponds to the time-averaged radial position where the density-weighted surface density is the closest to 250 g cm$^{-2}$, following the definition of \cite{Speicher2023MNRAS.526.1388S}. In the $a_*=0$ burst simulations, $r_{in}$ moves outwards from $\approx12.9$ km in time interval A to $\approx14.4$ km in interval I due to PR drag. In the $a_*=0.2$ burst simulations, $r_{in}\approx13$ km in time interval D and reaches $\approx13.5$ km in interval I. In the no-burst simulations, $r_{in}$ is $\approx12.9$ km ($a_*=0$) and $\approx11.5$ km ($a_*=0.2$), both close to the respective radii of the innermost stable circular orbits $r_\mathrm{isco}$ of $12.85$ km ($a_*=0$) and $11.41$ km ($a_*=0.2$). The outer radius $r_{out}$ is set to 300 km, close to the outer simulation boundary but not so close as to introduce boundary effects.%, $G$ the gravitational constant, and $k$ is the Boltzmann constant.
% We place the inner radius $r_{in}$ at the radial position where the surface density is the closest to $2.5\times10^2$ g cm$^{-2}$, time-averaged over a respective time interval. The resolution of the simulation decreases beyond 50 km, so we choose $r_{out}=50$ km. 
% The distance $D_L$ between the simulated neutron star system and a fictional observer is set to 6 kpc.

% We assume an inclination angle of $\cos i =1$. The redshift $1+z$ is \citep[][]{Hanawa1989ApJ...341..948H}
% \begin{equation}
%     1+z = \left(1+\frac{3MG}{c^2 r}\right)^{-1/2}\label{eq:oneplusz}.
% \end{equation}

The emitted spectrum will differ from a blackbody due to scattering processes. These deviations are captured with a color correction factor $f_{c}$, \citep[][]{Done2012MNRAS.420.1848D},
\begin{equation}
%      f_{c}\left ( T_{eff} \right ) = \left\{\begin{matrix}1 & T_{eff} < 3\times 10^4\,\mathrm{K} \\
% \left ( \frac{T_{eff}}{3\times 10^4\,\mathrm{K}} \right )^{0.82} &3\times 10^4\,\mathrm{K}\leq  T_{eff} \lesssim 10^5\,\mathrm{K} \\
% \left ( \frac{72\,\mathrm{keV}}{kT_{eff}} \right )^{1/9}  & T_{eff}\gtrsim 10^5\,\mathrm{K}  \\\end{matrix}\right.,
 f_{c}\left ( T_{eff} \right )  = \left ( \frac{72\,\mathrm{keV}}{kT_{eff}} \right )^{1/9} .
\label{eq:fc}
\end{equation}
%which depends on the effective temperature $T_{eff}$ \pcf{You've been using $T_{eff}$ since eq. (2), so you should probably define it earlier than here.}. 
In our calculations, $f_c$ varies between $\approx1.5$ and $\approx2.2$, consistent with previous results \citep[e.g.,][]{Kubota2010ApJ...714..860K,Suleimanov2012A&A...545A.120S}. Color corrections $>1$ will shift the emission to higher energies. 

\begin{figure}
    \centering
    \includegraphics[width = 0.45 \textwidth]{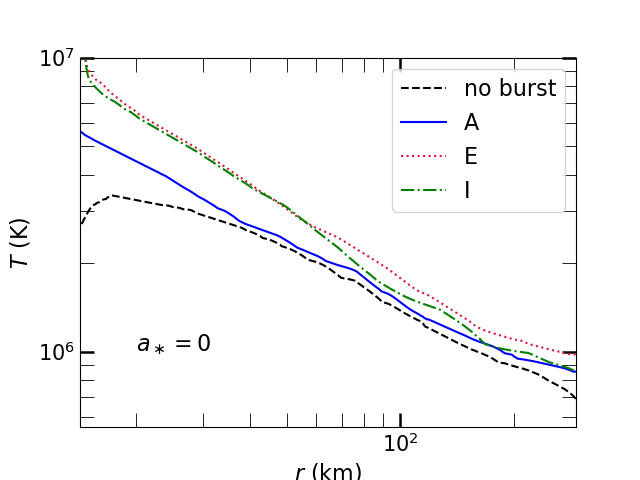}
    \caption{Radial temperature profile of the $a_*=0$ burst simulations for intervals A (blue solid line), E (red dotted line), and F (green dash-dotted line). The temperature profile of the no-burst simulation (black dashed line) is from interval F. The temperatures are located where the optical depth due to Thomson scattering, integrated towards the equatorial plane, first surpasses 1 and are equated to $T_{eff}$ for eq. \ref{eq:nuFnu} and \ref{eq:fc}. The plotted temperature profiles have been smoothed for visualization by calculating moving averages with varying window sizes.}
    \label{fig:radiusTemperature}
\end{figure}

With the assumption that the accretion disk radiates as a color-corrected
blackbody at every radius, we equate $T_{eff}$ to the gas temperature where $\tau$ is first $\geq1$, with $\tau$ being the optical depth due to Thomson scattering integrated towards the equatorial plane. Examples of radial temperature profiles for time intervals A (blue solid line), E (red dotted line) and I (green dash-dotted line) are shown in Fig. \ref{fig:radiusTemperature} for the $a_*=0$ burst simulation. For visualization, the plotted temperature profiles have been smoothed by calculating the moving average with varying window sizes. Due to the strong illumination by the X-ray burst, the disk temperature is higher during the burst than in interval F of the no-burst simulation (black dashed line). The disk remains hot in the burst tail, visible by the little difference in temperature profiles of intervals I and E. The disk temperatures are also spin dependent, with the simulated $a_*=0.2$ disk being hotter than the $a_*=0$ disk \citep[see also][]{Speicher2023MNRAS.526.1388S}. We again use linear interpolation of the simulation data and $dr=10$~m to find the $T_{eff}$ values for use in Eqs.~\ref{eq:nuFnu} and~\ref{eq:fc}.

% For all time intervals for the $a_*=0$ burst simulation, time intervals past the red dashed line in Fig. \ref{fig:lightcurve} for the $a_*=0.2$ burst simulation, and for the time interval marked with the slanted line in Fig. \ref{fig:lightcurve} for the no-burst simulations, w
With our interpolated values, we numerically integrate eq. \ref{eq:nuFnu} using the trapezoidal method for a linearly spaced, 600-entry-long energy grid between $\approx8\times10^{-3}$ keV and $\approx33$ keV. For the $a_*=0$ burst simulation, we calculate spectra for all time intervals A-I. The $a_*=0.2$ burst simulation experienced a transient numerical event at $\lesssim 1$ s \citep[see also][]{Speicher2023MNRAS.526.1388S}, so we only calculate spectra for intervals D-I (those past the red dashed line in Fig. \ref{fig:lightcurve}). We compare each spectrum from the burst simulations with the spectra of the corresponding no-burst simulations at their time interval F.

\begin{figure*}
    \centering
\includegraphics[width = 0.9 \textwidth]{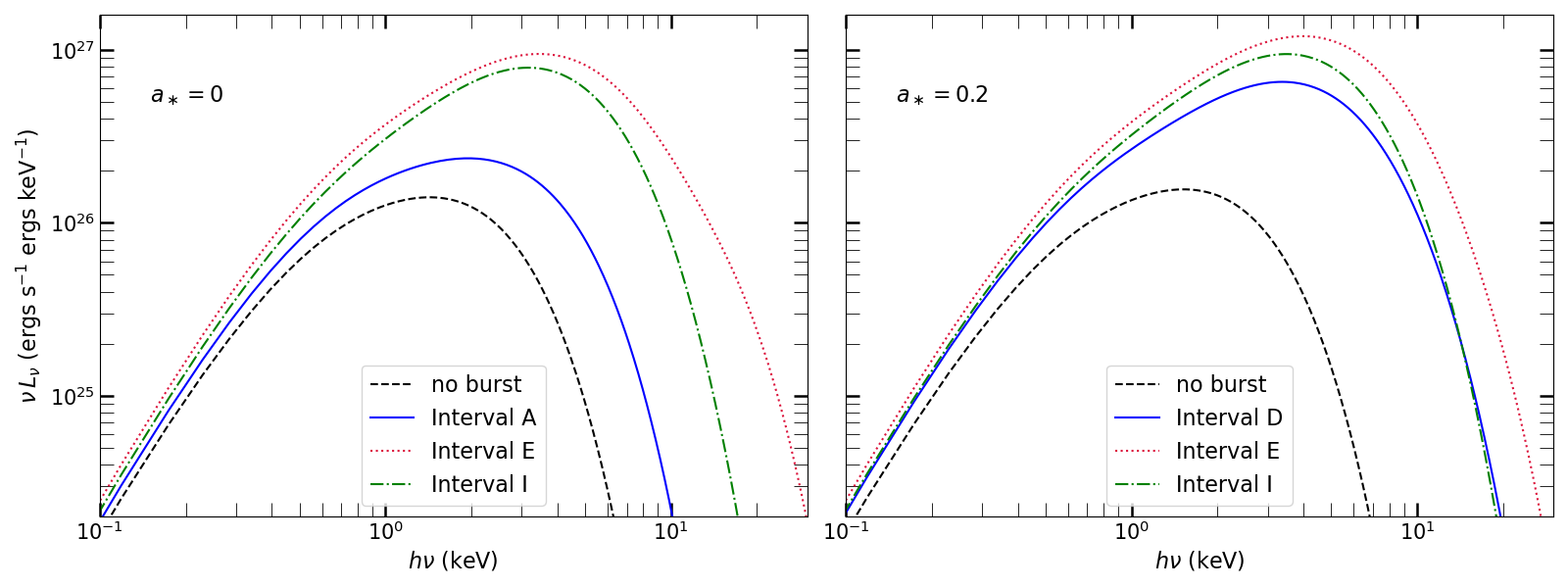}
    \caption{Left panel: Radially integrated disk spectra of the $a_*=0$ simulations. The spectra are integrated between the respective inner disk radius and 300 km (eq. \ref{eq:nuFnu}). The time intervals and line styles are the same as in Fig. \ref{fig:radiusTemperature}. Right panel: Same as the left panel, but for the $a_*=0.2$ simulations, and the spectra of time interval D instead of A being shown. }
    \label{fig:bothSpectra}
\end{figure*}
Fig. \ref{fig:bothSpectra} shows integrated spectra (eq. \ref{eq:nuFnu}) for time intervals A (blue solid line), E (red dotted line), and I (green dash-dotted line) for the $a_*=0$ burst simulation (left panel) and for time intervals D (blue solid line), E (red dotted line), and I (green dash-dotted line) for the $a_*=0.2$ burst simulation (right panel). 
%except for the burst rise interval of the spin simulation (blue solid line, right panel). While the burst rise interval is centered at $\bar{t}\approx0.6$ s for the no-spin simulation, it is centered at $\bar{t}\approx1.8$ s for the spin simulation (first interval past the red dashed line in Fig. \ref{fig:lightcurve}). 
Compared to the no-burst spectra of interval F (black solid lines), the disk spectra of both the $a_*=0$ and $a_*=0.2$ burst simulations are enhanced in magnitude and peak at higher energies. The enhanced magnitude and shift to higher energies is still present for the spectra of interval I due to the maintained high disk temperatures (Fig. \ref{fig:radiusTemperature}). Furthermore, for a respective time interval, the $a_*=0.2$ spectrum has a greater amplitude than the $a_*=0$ spectrum due to the higher disk temperatures in the $a_*=0.2$ simulation.

%Due to the disk heating (Fig. \ref{fig:radiusTemperature}) and $\menhance$ (eq. \ref{eq:menh}, Fig. \ref{fig:menh}),
% are the same of the the burst peak (red dotted line), and the burst tail (green dash-dotted line) are centered at $\bar{t}\approx2.2$ s, and $\bar{t}\approx3.9$ s. The time interval of the no-burst spectrum (black dashed line) is centered at $\bar{t}\approx2.6$ s for both panels. The time intervals used for the burst rise interval (blue solid line) differ 
% disk emission of the no-spin burst simulation in the burst rise (blue solid line), 
% , corresponding to the time intervals centered at $\bar{t}\approx0.6$ s, $\bar{t}\approx2.2$ s, and $\bar{t}\approx3.9$ s. The no-burst ( spectrum uses simulation data 

% The right panel of Fig. \ref{fig:bothSpectra} shows the disk spectra for the spin simulation. Like in the no-spin case, the spectra for the burst peak and the burst tail is calculated for the time intervals centered at $\bar{t}\approx2.2$ s and $\bar{t}\approx3.9$ s respectively. The burst peak time interval is centered at $\bar{t}\approx1.8$ s. 
% Similar to the emission from the no-spin burst simulation, the emission of the spin simulation is enhanced and shifted towards higher energies during the burst. In the burst rise at $\bar{t}\approx1.8$ s, the spectra peaks at $\approx1.5$ keV higher energies than for the no-burst simulation. Compared to the no-burst simulation, the spectral peak is at $\approx2$ keV higher energies at the burst peak and at $\approx1.7$ keV higher energies in the burst tail.
\subsection{Definition of \texorpdfstring{$f_a$}{} and the mass accretion rate enhancement}\label{subsec:definefamenh}

\begin{figure*}
    \centering
    \includegraphics[width = 0.9 \textwidth]{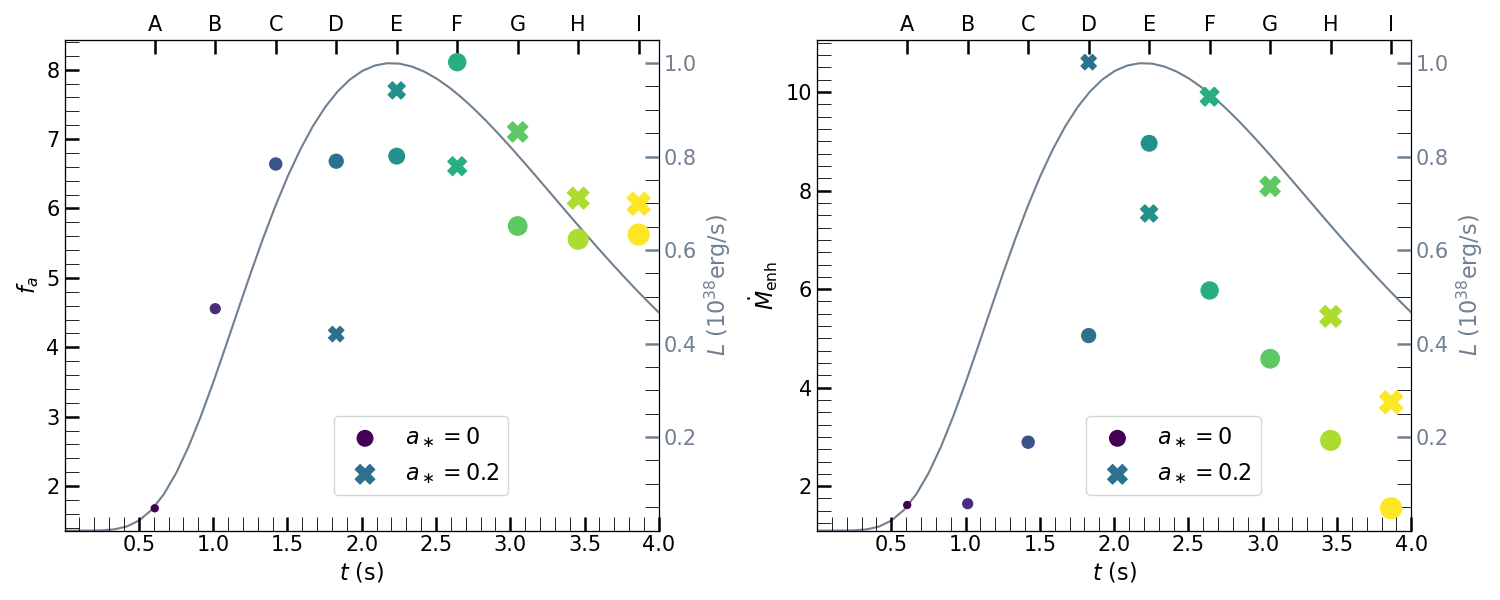}
    \caption{Left panel: The evolution of the normalization factor $f_a$, overplotted with the burst lightcurve (eq. \ref{eq:luminosity}, same as in Fig.\ref{fig:lightcurve}) for the $a_*=0$ (circle markers) and $a_*=0.2$ (cross markers) burst simulations. The size of the points in both panels increase with time. Their colors correspond to the shadings in Fig.\ref{fig:lightcurve} for their respective time interval. With some scatter, $f_a$ follows the burst lightcurve for both simulations. Right panel: Same as the left panel, but showing the evolution of the mass accretion rate enhancement. Due to PR drag during the burst, $\menhance$ calculated at $r_\mathrm{isco}$ (eq. \ref{eq:menh}) increases in the $a_*=0$ burst simulation (circle markers) from $\approx1.6$ to $\approx9$ and decreases again to $\approx1.6$. For the $a_*=0.2$ burst simulation (cross markers), $\menhance$ is $\approx10.6$ in interval F, and decreases to $\approx3.7$ in interval I.}
    \label{fig:famenh}
\end{figure*}

To quantify the emission enhancement of the calculated spectra (eq. \ref{eq:nuFnu}, Fig. \ref{fig:bothSpectra}), we utilize the normalization factor $f_a$. We calculate $f_a$ as the ratio between the maximum emission of the burst and the no-burst simulation\footnote{In spectral analysis, $f_a$ is defined as the scaling factor of the preburst emission \citep[e.g.,][]{Worpel2013ApJ...772...94W}, while eq.\ref{eq:fa} measures how the disk spectrum amplitude changes during the burst. Both definitions track the enhancement of the persistent emission and will evolve similarly.},
\begin{equation}
    f_a = \frac{\left(\nu L_\nu\right)_{\mathrm{max,\, b}}}{\left(\nu L_\nu\right)_{\mathrm{max,\,nb}}}.\label{eq:fa}
\end{equation}
The left panel of Fig. \ref{fig:famenh} shows the evolution of $f_a$ (eq. \ref{eq:fa}) for the $a_*=0$ (circle markers) and the $a_*=0.2$ (cross markers) burst simulation as a function of time. The marker size increases with time and the colors correspond to the time interval shadings in Fig. \ref{fig:lightcurve}. The overplotted burst lightcurve (gray solid line, eq.\ref{eq:luminosity}) shows that $f_a$ follows the burst evolution with some scatter for both spin configurations. For the $a_*=0$ simulation, $f_a$ is $\approx1.7$ at interval A, reaches a maximum of $\approx8.1$ at interval F, and is $\approx5.6$ in interval I. In the $a_*=0.2$ burst simulation, $f_a\approx4.2$ in time interval D, reaches a maximum of $\approx7.7$ in interval E, and decreases to $\approx6.1$ in interval I.

During the burst, the mass accretion rate also increases due to PR drag \citep[][]{Fragile2020NatAs...4..541F,Speicher2023MNRAS.526.1388S} We calculate the mass accretion rate enhancement $\menhance$ as the ratio of the time-averaged mass accretion rates between the burst and the no-burst simulations, measured at the respective $r_\mathrm{isco}$,
\begin{equation}
    \menhance = \left.\frac{\left<\dot{M}_{\mathrm{b}}\right>}{\left<\dot{M}_{\mathrm{nb}}\right>}\right\vert_{r= r_\mathrm{isco}}\label{eq:menh}.
\end{equation}
The right panel of Fig. \ref{fig:famenh} shows the evolution of $\menhance$. The size and colors of the scatter points are the same as for the left panel. For the $a_*=0$ burst simulation, $\menhance$ is initially $\approx1.6$ at interval A, reaches $\approx9$ at the burst peak at interval E, and decreases to $\approx1.6$ in the burst tail at interval I. In the $a_*=0.2$ burst simulation, the first considered time interval D has the highest $\menhance$ of $\approx10.6$, decreasing slightly to $\approx7.5$ at interval E and reaching $\approx3.7$ in time interval I.

\section{Results}\label{sec:results}

While the increase in the persistent emission captured by $f_a$ has been proposed to be connected to the $\menhance$ caused by PR drag, how well $f_a$ and $\menhance$ correlate remains unclear. In section \ref{subsec:favsmdot}, we examine the connection between $f_a$ and $\menhance$. Because $f_a$ has also been used to account for the soft excess, in Section \ref{subsec:impactSoftExcess} we investigate the contribution of the persistent emission to the soft excess.

\subsection{Emission and mass accretion rate relationship}\label{subsec:favsmdot}

% \begin{figure}
%     \centering
%     \includegraphics[width = 0.45 \textwidth]{pictures/faBoth.png}
%     \caption{The evolution of the normalization factor $f_a$, overplotted with the burst lightcurve (eq.\ref{eq:luminosity}, same as in Fig. \ref{fig:lightcurve}) for the no-spin (circle markers) and the spin simulation (diamond markers). The size of the scatter points in both panels increase with time. Their colors correspond to the shadings in Fig. \ref{fig:lightcurve} for their respective time interval. With some scatter, $f_a$ follows the burst lightcurve for both simulations.}
%     \label{fig:favslb}
% \end{figure}
% Compared to the disk emission from the no-burst simulation (black dashed line), the emission of the burst simulation is enhanced. The emission enhancement is due to an increase in the mass accretion rate and disk heating in the burst simulation and can be captured with the normalization constant $f_a$. We define $f_a$ as the ratio between the burst simulation's peak emission and the no-burst simulation's peak emission (Fig. \ref{fig:wholeDiskSpectraStatic}, black dashed line). 

\begin{figure*}
    \includegraphics[width = 0.9 \textwidth]{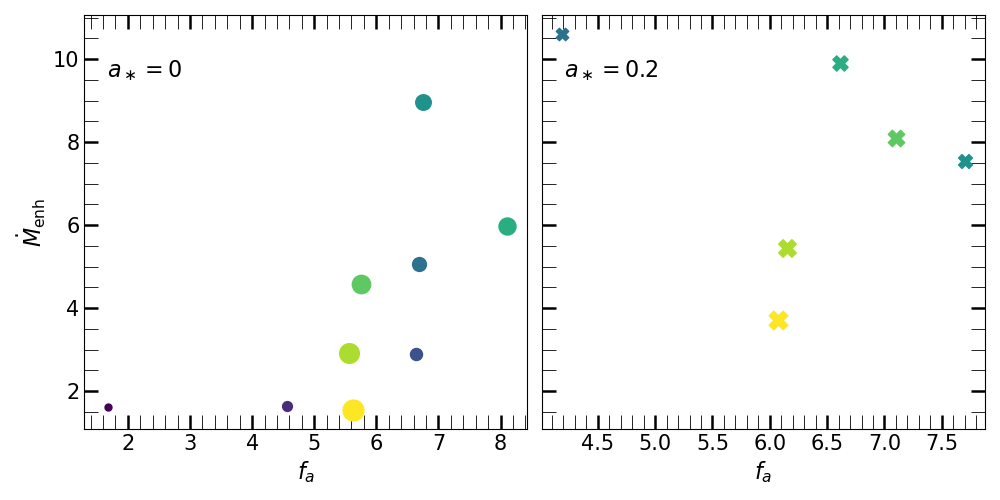}
    \caption{Left panel: Plot of $f_a$ versus $\menhance$ for the $a_*=0$ simulation. The colors are the same as in Fig. \ref{fig:lightcurve} and the marker sizes increase with time. $f_a$ and $\menhance$ are weakly linearly correlated, with a Pearson coefficient of $0.63$. Right panel: Same as the left panel, but for the $a_*=0.2$ simulation. Here, scatter affects the $f_a$ versus $\menhance$ relationship more significantly than for the $a_*=0$ simulation. The correlation between $\menhance$ and $f_a$ is only positive when restricting the time intervals E-I, with a correlation factor of $0.56$.}
    \label{fig:favsmenh}
\end{figure*}

The relationship of $f_a$ versus $\menhance$ for the $a_*=0$ (left panel) and the $a_*=0.2$ (right panel) burst simulations can be seen in Fig. \ref{fig:favsmenh}. For the $a_*=0$ simulation, a linear fit yields $\menhance = \left(0.9\pm0.4\right)f_a + \left(-1.0\pm2.4\right)$ and a Pearson correlation coefficient of $0.63$. Hence, $f_a$ and $\menhance$ are weakly related for the $a_*=0$ simulation.

Scatter affects the $f_a$ versus $\menhance$ relationship significantly more in the $a_*=0.2$ simulation (Fig. \ref{fig:favsmenh}, right panel). Fitting a linear function yields $\menhance = \left(-0.6\pm1.1\right)f_a + \left(11.1\pm6.7\right)$ and a Pearson correlation coefficient of $-0.26$. The negative correlation is due to the large $\menhance$ despite a low $f_a$ in interval D. Ignoring this time interval and using only intervals E-I for fitting gives $\menhance = \left(2.0\pm1.7\right)f_a + \left(-6.2\pm11.4\right)$ and increases the Pearson correlation coefficient to $0.56$. For the $a_*=0$ burst simulation, restricting the fit to intervals E-I yields $\menhance = \left(1.6\pm1.2\right)f_a + \left(-5.5\pm7.7\right)$ and reduces the Pearson correlation coefficient slightly to $0.62$. Neglecting the burst rise, the slopes of the $f_a$ versus $\menhance$ relationships of the $a_*=0$ and $a_*=0.2$ simulations agree with each other within their uncertainties. However, the large uncertainties in slopes and the relatively low Pearson correlation coefficients ($\leq0.63$) point towards a rather weak $f_a$ versus $\menhance$ relationship. 

The weak $f_a$ versus $\menhance$ stems largely from the high temperatures in the burst tail, which decreases $f_a$ less than $\menhance$ (Fig. \ref{fig:famenh}). In the $a_*=0$ simulation, $f_a$ increases from interval A to E by a factor of $4.02$ while $\menhance$ increases by a factor of $5.54$. However, as $\menhance$ decreases significantly again by a factor of $5.78$ between interval E and I, $f_a$ decreases only by a factor of $1.20$. Considering only intervals A-E of the $a_*=0$ simulation yields a correlation coefficient of $0.73$. Hence, $f_a$ and $\menhance$ are more correlated in the burst rise, and $f_a$ becomes more decoupled from the $\menhance$ evolution in the burst tail. Since $f_a$ follows the high disk temperatures in the burst tail rather than $\menhance$, $f_a$ correlates with temperature. %$the high disk temperatures in the burst tail cause $f_a$ to diverge from the $
%\menhance$ evolution, $f_a$ and the disk temperatures are correlated \pcf{The logic in this sentence is flawed. You claim the temperature causes $f_a$ to diverge and then use this as evidence that they are correlated. But if one thing is the cause of the other, then by definition, they must be correlated.}. 
Between $f_a$ and the disk temperature at $r_{in}$, we find a high Pearson correlation factor of $0.93$ for the $a_*=0$ simulation (based on intervals A-I) and $0.86$ for the $a_*=0.2$ simulation (based on intervals E-I). Because $f_a$ traces the disk temperature evolution, which especially in the burst tail does not represent $\menhance$, $f_a$ is not a reliable tracer of $\menhance$ in observations.

\begin{figure*}
    \centering
    \includegraphics[width = 0.9 \textwidth]{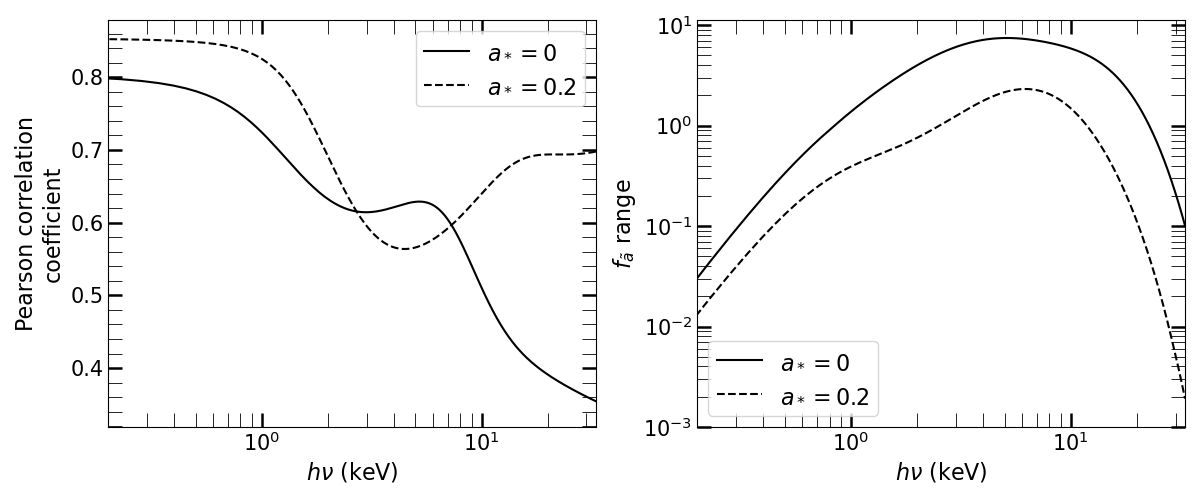}
    \caption{Left panel: Pearson correlation coefficient between $f_{\Tilde{a}}$ (eq. \ref{eq:faVary}) and $\menhance$ (eq. \ref{eq:menh}) as a function of energy. For both the $a_*=0$ (solid line) and $a_*=0.2$ (dashed line) simulations, $f_{\Tilde{a}}$ correlates most strongly with $\menhance$ at energies below 1 keV. Right panel: The range of $f_{\Tilde{a}}$ for the $a_*=0$ (solid line) and the $a_*=0.2$ (dashed line) simulations. With decreasing energy and beyond $\approx10$ keV, $f_{\Tilde{a}}$ varies less. 
    %\pcf{It also varies less at high energies, so your statement depends on where you start looking.}
    }\label{fig:otherFas}
\end{figure*}

The weak $f_a$ versus $\menhance$ correlation changes with a different definition of the normalization factor (eq. \ref{eq:fa}). At energy $h\nu$, the redefined normalization factor called $f_{\Tilde{a}}$ is the ratio between the emission at that energy of the burst simulation and the maximum emission of the no-burst simulation,
\begin{equation}
    f_{\Tilde{a}}(h\nu) = \frac{\left[\nu L_\nu (h\nu)\right]_{\mathrm{ b}}}{\left[\nu L_\nu\right]_{\mathrm{max,\,nb}}}.\label{eq:faVary}
\end{equation}
The left panel of Fig. \ref{fig:otherFas} shows the Pearson correlation coefficients between $f_{\Tilde{a}}$ and $\menhance$ at different energies. The correlation coefficient is based on time intervals A-I for the $a_*=0$ burst simulation (solid line). For the $a_*=0.2$ burst simulation (dashed line), only intervals E-I are used because the inclusion of interval D yielded significantly lower coefficients $\leq0.36$. The correlation coefficients are highest at lower energies, reaching values up to $0.80$ for $a_*=0$ and $ 0.85$ for $a_*=0.2$. However, the range in $f_{\Tilde{a}}$ decreases with decreasing energy for $h\nu \lesssim 10$ keV (Fig. \ref{fig:otherFas}, right panel). For example, at $0.51$ keV $f_{\Tilde{a}}$ ranges between $0.58$ and $0.92$ for the $a_*=0$ simulation and between $0.71$ and $0.84$ for the $a_*=0.2$ simulation. 
The small range and values in $f_{\Tilde{a}}$ at low energies differ from the $f_a$ inferred in observations \citep[e.g.,][]{Worpel2015ApJ...801...60W,Bult2022ApJ...940...81B,Guver2022MNRAS.510.1577G}, so the low energy $f_{\Tilde{a}}$ are not a better representation of the observational $f_a$ compared to eq. \ref{eq:fa}. Hence, finding a high correlation between the low energy $f_{\Tilde{a}}$ and $\menhance$ does not indicate that the observational $f_a$ factors would also have a high correlation as well. %, which has not been captured with our primary definition of the normalization factor (eq. \ref{eq:fa}). %Since the low-energy $f_{\Tilde{a}}$ have different properties than the observational $f_a$ factors, the observational $f_a$ factors are not expected to have equally high Pearson correlation coefficients as the low-energy $f_{\Tilde{a}}$. 
Obtaining these high coefficients in observational fits would require a redefinition of $f_a$. However, while $f_{\Tilde{a}}$ correlates well with $\menhance$ at small energies, the small variation of $f_{\Tilde{a}}$ at those energies might be challenging to measure in observations, especially if reflected and coronal emissions also extend to low energies (see section \ref{sec:discussion}).

\subsection{Impact of the spectral evolution on the soft excess}\label{subsec:impactSoftExcess}
In addition to the emission enhancement, burst-induced disk heating shifts the calculated disk emission (Fig. \ref{fig:bothSpectra}) to higher energies. For the $a_*=0$ burst simulation, the spectral peak occurs at $\approx0.5$ keV higher energy in time interval A than for the no-burst simulation spectrum. The spectral peak shift reaches $\approx2.9$ keV in interval F (its maximum) and decreases to $\approx1.7$ keV in interval I. The spectral peak is shifted similarly in the $a_*=0.2$ burst simulation. For time interval D, the spectral peak shift compared to its no-burst spectrum is $\approx1.9$ keV, reaching its maximum of $\approx2.5$ keV in interval E, and decreasing to $\approx2.0$ keV in time interval I.
% reacheof the no-spin burst simulation occurs at $\approx0.5$ keV higher energies in the burst rise. In the burst peak, the peak has shifted by $\approx1.7$ keV as compared to the no-burst emission. In the burst tail, the spectral peak only slightly shifted to lower energies, and occurs at $\approx1.5$ keV higher energies than for the no-burst simulation. The heating processes maintaining the high disk temperatures are discussed in section \ref{subsec:heatCool}. Because $f_a$ does not consider the location of the spectral peak, the normalization factor cannot account for the change in spectral shape.

\begin{figure*}
    \centering
    \includegraphics[width = 0.8 \textwidth]{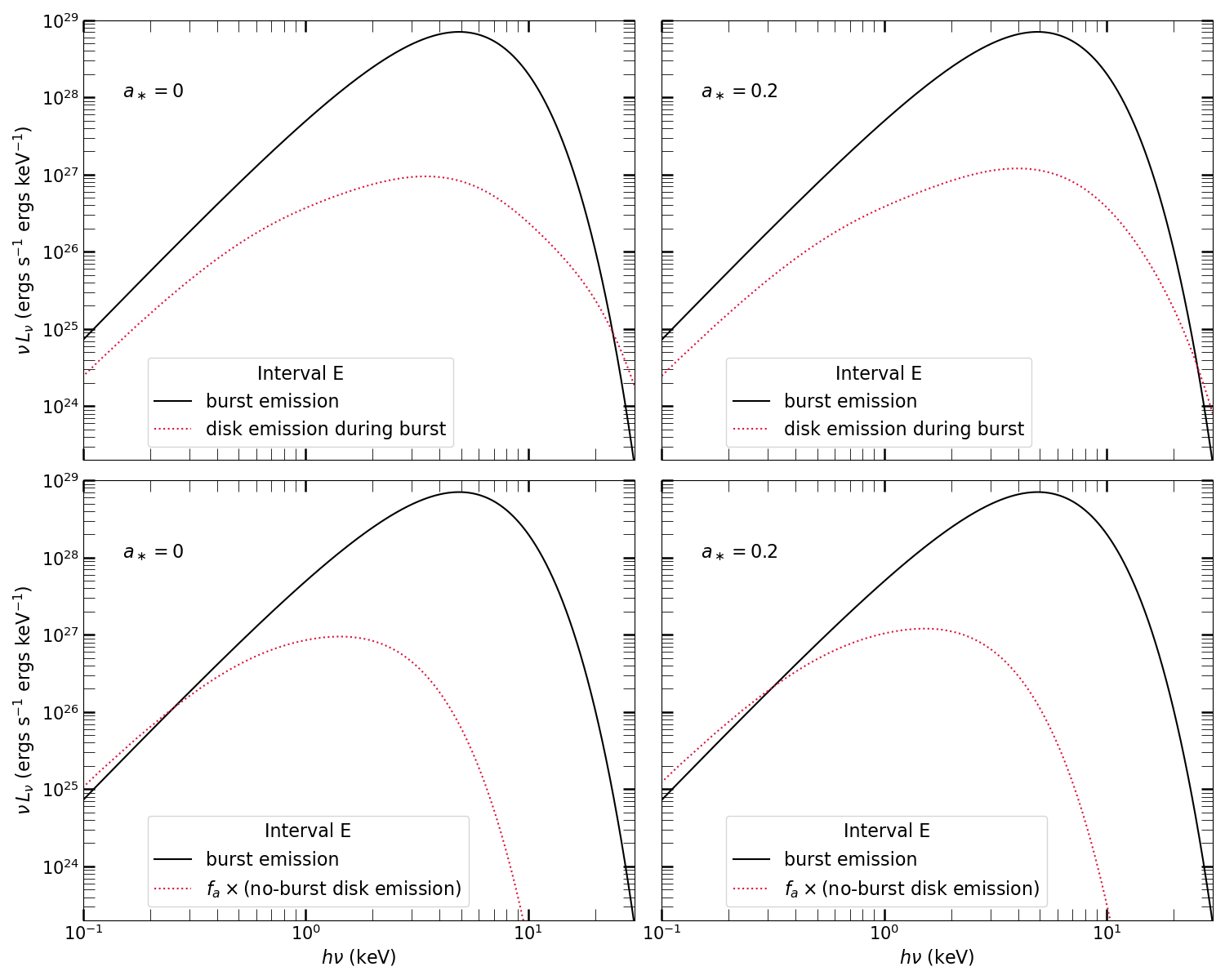}
    \caption{Top row: The disk spectra (red dotted lines, same as in Fig. \ref{fig:bothSpectra}) do not surpass the burst spectra (black solid lines) in time interval E for either the $a_*=0$ (left column) nor $a_*=0.2$ (right column) simulation, and thus do not contribute to a soft excess. Bottom row: Same as the top row, but instead of true disk spectra, no-burst spectra multiplied by $f_a$ are plotted. The scaled no-burst spectra exceed the burst spectra at low energies and generate an apparent soft excess. The disk spectra shifting to higher energies (top row) compared to them staying fixed in shape (bottom row) reduces their contribution to the soft excess.}\label{fig:softExcess300}
\end{figure*}

% Hence, scaling the no-burst emission with $f_a$ does not reflect the actual spectral evolution, with consequences on the soft emission of the accretion disk spectrum. Due to the spectral shift to higher energies, the calculated disk emission hardly produces a soft excess during the burst,
The shift of the disk emission to higher energies impacts its contribution to the soft excess, as shown in Fig. \ref{fig:softExcess300}. Here, we define the soft excess as the disk emission surpassing the burst emission at energies $\lesssim3$ keV. 
% The percentage of soft excess $\mathcal{S}$ can be quantified as \citep[see also][]{Speicher2022MNRAS.509.1736S},
% \begin{equation}
%     \mathcal{S} = \frac{\int^{3\,\mathrm{keV}}_{0.1\,\mathrm{keV}} \left[L_\mathrm{disk}(E) - L_\mathrm{burst}(E)\right]\,\mathrm{d}E}{\int^{3\,\mathrm{keV}}_{0.1\,\mathrm{keV}} L_\mathrm{disk}(E)\,\mathrm{d}E}\,\times100,\label{eq:softExcess}
% \end{equation}
% where $L_\mathrm{disk}$ and $L_\mathrm{burst}$ are the disk and burst emission respectively at a given energy $E$.
To calculate the burst emission, we assume it follows a blackbody with a blackbody temperature 
\begin{equation}
    T_{b}(t) = \left(\frac{L(t)}{4\pi \sigma_b R^2}\right)^{1/4}\left(1+z_*\right)^{-1}, 
\end{equation}
where $\sigma_b$ is the Stefan–Boltzmann constant. The gravitational redshift $1+z_*$ is \citep[][]{Lewin1993SSRv...62..223L},
\begin{equation}
    1+z_*  = \left(1-\frac{2GM}{Rc^2}\right)^{-1/2},
\end{equation}
with $G$ being the gravitational constant. The burst spectrum is evaluated with the blackbody function, 
\begin{equation}
   B_\nu(T)= \frac{2h\nu^3}{c^2}\frac{1}{ \exp\left [ h\nu/kT_b \right ]-1}.
\end{equation}

Assuming the neutron star emits isotropically,  the burst luminosity is $B_\nu$ multiplied by the redshifted neutron star surface area and $\pi$ \citep{Rybicki1986rpa..book.....R,Lewin1993SSRv...62..223L}:
\begin{equation}
    L_{\nu,burst} = \left(\pi B_\nu(T)\right)\left[4\pi\left(1+z_*\right)^{2} R^2\right].
\end{equation}
Note that these choices preserve the equality $4\pi\left(1+z_*\right)^{2} R^2 \sigma_b T_b^4 = 4\pi\left(1+z_*\right)^{2} R^2\,\pi\int B_\nu\,d\nu$. 

The top row of Fig. \ref{fig:softExcess300} shows the disk spectra for time interval E (red dotted lines, same as in Fig. \ref{fig:bothSpectra}) and the burst spectra (black solid lines) for the $a_*=0$ (left column) and $a_*=0.2$ (right column) simulations. Neither in time interval E nor during any other time interval does the disk emission surpass the burst emission at low energies, amounting to no contribution to a soft excess. Since we terminate the radial integration of the disk emission (eq. \ref{eq:nuFnu}) at 300 km due to the outer simulation boundary, the shown spectra do not include the low-temperature emission of further out accretion disk regions. Including those would raise the emission at low energies and may generate a soft excess. However, the disk region beyond 300 km will not be significantly impacted by the burst radiation, so its luminosity will likely remain low and constant during the burst. Hence, even with an extension beyond 300 km, the disk emission is not expected to produce a strong soft excess.

%\pcf{I don't understand this paragraph. We just said in the previous paragraph that there is no soft excess. But now in this paragraph there is a soft excess. If I understand the caption to Fig. 7 correctly, what you intend to say here is that *if you don't account for the change in shape of the disk emission during the burst*, then you would expect a soft excess, but that's not clear from the way this paragraph is written.} 
The absence of a soft soft excess during the burst simulations is due to the evolution of the accretion disk spectra. The bottom row of Fig. \ref{fig:softExcess300} shows the no-burst disk spectra of interval F (red dashed lines) multiplied by the respective $f_a$ of time interval E. At energies $\lesssim0.3$ keV, the scaled no-burst emission exceeds the burst emission (black solid lines) for both the $a_*=0$ (left panel) and $a_*=0.2$ (right panel) configurations, creating a soft excess. The spectral shape evolution to higher energies therefore reduces the accretion disk contribution to the soft excess.

To compare the soft excess from the disk emission to the soft excess due to reflection \citep{Speicher2022MNRAS.509.1736S}, we calculate the percentage of soft excess below 3 keV as \citep{Speicher2022MNRAS.509.1736S} 
\begin{equation}
    \mathcal{S} = \frac{\int_{0.01\,\mathrm{keV}}^{3\,\mathrm{keV}}\left[F_s(E) - F_{bb}(E)\right]\,dE}{\int_{0.01\,\mathrm{keV}}^{3\,\mathrm{keV}} F_s(E)\,dE}\times 100,\label{eq:sPer}
\end{equation}
with $F_s(E)$ being the disk emission and $F_{bb}(E)$ the burst emission as a function of energy $E$. The disk emission of the burst simulations (Fig. \ref{fig:softExcess300}, top row) does not surpass the burst emission below 3 keV, yielding no positive $\mathcal{S}$ for any time intervals or spin configurations. 
%\pcf{How can this be? The above expression would only be 0 if $F_s(E) = F_{bb}(E)$ for all $E$, which is clearly not the case. Do yoou mean that you only include contributions where $F_s > F_{bb}$? If so, then that's not clear.}. 
For the scaled no-burst emission (Fig. \ref{fig:softExcess300}, bottom row), we exclude instances where $F_s(E) - F_{bb}(E) < 0$ when evaluating the integral in the numerator of eq. \ref{eq:sPer}. Otherwise, $\mathcal{S}$ would also never be positive. Despite this modified calculation of $\mathcal{S}$, the percentage of soft excess is $\lesssim0.2\%$ for all time intervals and spin configurations and thus negligible. Considering the strong soft excess due to reflection of up to $\approx38\%$ found by \cite{Speicher2022MNRAS.509.1736S}, calculated using comparable simulation data \citep[see also ][]{Fragile2020NatAs...4..541F}, these results imply that disk reprocessing is the main contributor to the soft excess.

% The left panel of Fig. \ref{fig:softExcess300} shows the burst (black lines) and the disk emission at the same time interval as in Fig. \ref{fig:wholeDiskSpectraStatic}. Burst and disk spectra of the same time interval are plotted with the same linestyle. The disk emission is shown for time intervals at the rise (blue solid lines), at the peak (red dotted lines), and at the tail of the burst (green dash-dotted lines). Because the spectra integrated up to 50 km (lines in left panel without stars, same as in Fig. \ref{fig:wholeDiskSpectraStatic}) do not yield a soft excess, we extend the integration of eq.\ref{eq:nuFnu} beyond the well-resolved region up to 300 km (lines with stars). However, the disk emission does not contribute to the soft excess even with the larger considered radial range (for a further discussion, see section \ref{subsec:obsImpl}).

% The right panel of Fig. \ref{fig:softExcess300} shows the spectrum calculated from the no-burst simulation and radially integrated up to 300 km, multiplied with the $f_a$ factor of the corresponding time interval. The color coding and linestyles are the same as for the left panel. The scaled no-burst emission exceeds the burst emission at $\lesssim0.08$ keV in the burst rise (right panel, blue solid line), and at $\lesssim0.35$ keV peak of the burst (right panel, red dotted line) and in the tail of the burst (right panel, black dash-dotted line). The presence of a soft excess for the scaled no-burst emission highlights the impact of the change in the spectral shape of the disk emission during the burst.

\section{Discussion}\label{sec:discussion}
As varying the normalization of the accretion disk emission can improve the fit applied to burst observations, the $f_a$ factor has become a standard tool in the burst analysis repertoire. However, as shown in section \ref{sec:results}, $f_a$ carries limited physical meaning. The normalization factor cannot accurately constrain $\menhance$ due to the relatively weak correlation of $f_a$ and $\menhance$. However, additional factors may further impact the correlation during real X-ray bursts, and are discussed in this section.%Moreover, additional factors not considered here could affect $f_a$ further in real neutron star systems. In this section, we discuss how $f_a$ could be further impacted in observations.

% While $f_a$ and $\menhance$ are similarly correlated for the $a_*=0$ and $a_*=0.2$ simulation within their uncertainties when neglecting the burst rise, 
% The slope and correlation between $f_a$ versus $\menhance$ could differ for different accretion disks. Disk heating in the form of viscous heating will depend on the viscosity coefficient \citep[e.g.,][]{Fragile2018Instability}. A smaller viscosity coefficient causes less disk heating, which may strengthen the $f_a$ versus $\menhance$ correlation. Thus, the amount of heating and consequently the strength of the $f_a$ versus $\menhance$ correlation could vary for differently viscous disks. %impacting our calculated $f_a$ versus $\menhance$ relationships (section \ref{subsec:favsmdot}). Viscous heating is proportional to the viscosity coefficient \citep[e.g.,][]{Fragile2018Instability}}, and thus has a different efficiency for differently viscous disks. If viscous heating increased the disk temperature less, the disk spectrum will be less affected by disk heating, which could strengthen the correlation between $f_a$ and $\menhance$. 

Burst properties will affect the $f_a$ versus $\menhance$ correlation. The burst photon energies impact the extent of disk heating and depend on the abundances in the atmosphere of the neutron star and its surface gravity \citep[e.g.,][]{Suleimanov2011A&A...527A.139S,Nattila2015A&A...581A..83N}. Since disk heating affects $f_a$, its extent cannot only strengthen or weaken the $f_a$ vs $\menhance$ relationship, but also change its slope. Thus, neutron star systems can be expected to display a variety of $f_a$ vs $\menhance$ correlations,  which will complicate constraining $\menhance$ using $f_a$.

%An X-ray burst with a large flux but low burst photon energies could induce significant PR-drag and thus increase $\menhance$, while causing only a marginal increase of disk temperatures. Minor disk heating may strengthen the $f_a$ versus $\menhance$ correlation. However, a minor increase in disk temperatures and thus $f_a$ will also change the slope of the $f_a$ versus $\menhance$ relationship. Thus, burst properties could vary the correlation and slope of the $f_a$ versus $\menhance$ relationship, which will complicate constraining $\menhance$ with $f_a$.%while we found a rather weak relationship between $f_a$ and $\menhance$, the relationship could be stronger but also have a different slope for other neutron star systems, 

In addition, the spectral evolution of the corona will impact $f_a$. In this paper, we focused on the spectral evolution of the accretion disk. However, the persistent emission also encompasses the emission of the corona, which radiates at high X-ray energies when not irradiated by an X-ray burst. During X-ray bursts, burst photons cool the corona \citep[e.g.,][]{Fragile2018ApJ...867L..28F,Speicher2020MNRAS.499.4479S}, leading to a drop in the observed hard X-ray emission \citep[e.g.,][]{Maccarone2003A&A...399.1151M,Ji2014ApJ...782...40J,SanchezFernandez2020A&A...634A..58S}. As the coronal emission at high X-ray energies decreases, the  emission at low X-ray energies can increase depending on burst and coronal parameters \citep{Speicher2020MNRAS.499.4479S}. An increase in soft coronal emission will increase $f_a$. While coronal cooling is accompanied by an increase in the mass accretion rate \citep{Fragile2018ApJ...867L..28F}, the mass accretion rate evolution might be different in the corona compared to the accretion disk. Furthermore, the soft X-ray increase due to coronal cooling is not expected to trace the coronal accretion rate accurately because the emission increase depends on burst and coronal parameters. Hence, accounting for the coronal evolution during an X-ray burst will further cloud the $f_a$ versus $\menhance$ relationship.

While the abovementioned factors complicate the $f_a$ versus $\menhance$ correlation, a different heating description could potentially strengthen it. The weak correlation largely stems from the high disk temperatures in the burst tail. Despite the receding burst luminosity in the tail, the burst radiation can still efficiently heat the disk. The high heating efficiency may be maintained by the evolving disk properties, such as a decreasing optical depth in the disk, which allows the burst radiation to heat deeper in the disk. Longer burst simulations are needed to fully decipher the disk evolution in the burst tail. In addition, accounting for magnetic fields in the disk structure could impact the thermodynamics of the disk \citep[e.g.,][]{Sadowski2016MNRAS.459.4397S} and this may avert the high disk temperatures maintained in the burst tail found here and therefore strengthen any $f_a$ versus $\menhance$ relationship. Numerical simulations of accretion disks impacted by X-ray bursts that follow the disk response through to the end of the burst are needed to better understand the thermal evolution of the disk.

\section{Conclusions}\label{sec:conclusion}
% Furthermore, while $f_a$ increases mainly due to an increase in the mass accretion rate, the shape of the disk emission changes due to heating as well. The disk temperature evolution shifts the disk emission towards higher energies. While not being associated with disk heating, a change in the persistent emission shape has been inferred during some burst observations \citep[e.g.,][]{Degenaar2016MNRAS.456.4256D,Kajava2017MNRAS.472...78K,Yu2024A&A...683A..93Y}. To account for the disk heating, the persistent emission could be fitted with a multicolor blackbody that is not only scaled by an $f_a$ factor but also has a temperature varying as $T_{\mathrm{no\,burst}}+\Delta T$. However, while this modification could classify the disk emission more accurately, the normalization factor would still not yield a quantitative mass accretion rate enhancement for the reasons mentioned above. Still, neglecting the evolution in the emission shape could introduce biases, such as incorrectly ascribing parts of the observed emission to the accretion disk.

X-ray spectral observations of Type I X-ray bursts are often fit with the persistent emission varying in magnitude, quantified with the normalization factor $f_a$ \citep[e.g.,][]{Worpel2013ApJ...772...94W,Worpel2015ApJ...801...60W,Bult2022ApJ...940...81B}. In addition, $f_a$ has been used to account for an often-observed soft excess \citep[e.g.,][]{Keek2018ApJ...855L...4K,Guver2022ApJ...935..154G}. The increase of $f_a$ is usually attributed to an increase in the mass accretion rate due to PR drag. 

This paper examined the $f_a$ versus $\menhance$ relationship. With the simulation data of \cite{Speicher2023MNRAS.526.1388S}, we calculated the accretion disk emission during the burst evolution (section \ref{sec:methods}) and found that $f_a$ traces the disk temperature evolution but only shares a weak relationship with $\menhance$ (section \ref{subsec:favsmdot}), suggesting that $f_a$ cannot accurately capture $\menhance$ in observations (section \ref{sec:discussion}). 

% While assuming other disk models could strengthen the relationship, considering the impact of the burst radiation on other parts of the accretion flow and possible dependencies on burst and disk properties suggests that $f_a$ cannot accurately capture $\menhance$ in observations (section \ref{subsec:obsImpl}). 
% We found that the disk emission traces the burst lightcurve and the mass accretion rate enhancement for both the spin and the no-spin simulation (section \ref{sec:results}). However, we find that the $f_a$ versus $\menhance$ correlations are subject to significant uncertainties, for instance due to disk heating, and differ for the two simulation setups, which will complicate the $\menhance$ estimation from burst observations. 

Moreover, heating of the disk by the burst causes the disk emission to peak at increasingly higher energies, which reduces its contribution to the soft excess. The X-ray reflection spectra produced by the illuminated disk will therefore contribute more significantly to the soft excess than changes to the persistent emission (section \ref{subsec:impactSoftExcess}). Therefore, X-ray reflection models appropriate for X-ray bursts \citep[e.g.,][]{ball04,garcia22} should be used to account for any observed soft excesses found in the spectra of X-ray bursts.
%A change in the persistent emission shape has been inferred during some burst observations \citep[e.g.,][]{Degenaar2016MNRAS.456.4256D,Kajava2017MNRAS.472...78K,Yu2024A&A...683A..93Y}. 

To account for the spectral shape evolution, the disk emission could be fitted as a scaled multicolor blackbody with a varying temperature $T_{\mathrm{nb}}+\Delta T$, where $T_{\mathrm{nb}}$ is the temperature at $r_{in}$ with no burst \citep[assuming the disk emission is fitted with the \texttt{diskbb} model by][]{Makishima1986ApJ...308..635M}. Alternatively, a normalization factor focused solely on changes at low energy may give more insight into $\menhance$ (section \ref{subsec:favsmdot}). However, both approaches require very high quality datasets \citep[e.g.,][]{intZand2013A&A...553A..83I}, which will continue to be rare until the deployment of the next high throughput X-ray observatory. The current approach of scaling the persistent emission with $f_a$ is already pushing the current data quality to its limits and will give an inaccurate picture of $\menhance$, as well as incorrectly classifying parts of the observed emission as originating from the accretion disk. In its current form, $f_a$ may provide some insight into the temperature of disk during an X-ray burst, although the complicated structural changes of the disk will limit its utility to accurately probe the details of the disk. Since the burst-disk interaction leads to a rich range of physical processes ongoing simultaneously \citep[e.g.,][]{be05, Degenaar2018SSRv..214...15D, 2024Natur.627..763R}, spectral models motivated by the results of simulations such as these are likely the best strategy to infer the physical properties of the system.

\begin{acknowledgments}
JS acknowledges support from the
Department of Defense (DoD) through the National Defense Science
\& Engineering Graduate (NDSEG) Fellowship Program. DRB is supported in part by NASA award 80NSSC24K0212 and NSF grants AST-2307278 and AST-2407658. PCF gratefully acknowledges the support of NASA through award No 23-ATP23-0100.
\end{acknowledgments}

\bibliography{references}{}
\bibliographystyle{aasjournal}



\end{document}